\documentclass[twocolumn,showpacs,preprintnumbers,amsmath,amssymb,prl]{revtex4-1}

\usepackage{graphicx}% Include figure files
\usepackage{dcolumn}% Align table columns on decimal point
\usepackage{bm}% bold math
\usepackage{tabularx}% Table width
\usepackage{amsmath}% Higher math

\begin{document}

\title{Multiple nodeless superconducting gaps in noncentrosymmetric superconductor PbTaSe$_{2}$ with topological bulk nodal lines}

\author{M. X. Wang,$^1$ Y. Xu,$^1$ L. P. He,$^1$ J. Zhang,$^1$ X. C. Hong,$^1$ P. L. Cai,$^1$ Z. B. Wang,$^1$ J. K. Dong,$^1$ and S. Y. Li$^{1,2,*}$}

\affiliation{$^1$State Key Laboratory of Surface Physics, Department of
Physics, and Laboratory of Advanced Materials, Fudan University,
Shanghai 200433, China \\
$^2$Collaborative Innovation Center of Advanced Microstructures, Fudan University, Shanghai 200433, China}

\date{\today}

\begin{abstract}
Low-temperature thermal conductivity measurements were performed on single crystal of PbTaSe$_2$, a noncentrosymmetric superconductor with topological bulk nodal lines in the electronic band structure. It is found that the residual linear term $\kappa_0/T$ is negligible in zero magnetic field. Furthermore, the field dependence of $\kappa_0/T$ exhibits a clear ``$S$"-shape curve. These results suggest that PbTaSe$_2$ has multiple nodeless superconducting gaps. Therefore, the spin-triplet state with gap nodes does not play an important role in this noncentrosymmetric superconductor with strong spin-orbital coupling. The fully gapped superconducting state also meets the requirement of a topological superconductor, if PbTaSe$_2$ is indeed the case.
\end{abstract}

\pacs{74.20.Pq, 74.25.fc, 74.70.-b}

\maketitle

The search for exotic superconductors and elucidating their superconducting pairing mechanism is always an exciting field in condensed matter physics \cite{Norman}. In recent years, besides the iron-based superconductors \cite{XHChen}, the noncentrosymmetric superconductors and topological superconductors also attracted much attention \cite{Bauer,XLQi,Ando}. In a noncentrosymmetric superconductor, the lack of inversion symmetry introduces an antisymmetric spin-orbit coupling. This may split the electron bands by lifting the spin degeneracy, allowing admixture of spin-singlet and spin-triplet pairing states, thus line nodes in the superconducting gap \cite{Bauer,Frigeri}. The discovery of superconductivity in some topological materials has led to an enormous number of experimental and theoretical studies for topological superconductors (TSCs) \cite{XLQi,Ando}. The TSCs have a full pairing gap in the bulk and gapless surface states consisting of Majorana fermions \cite{XLQi}. The TSC is of great importance, since it is not only a new kind of exotic superconductor, but also one source of Majorana fermions for future applications in quantum computations \cite{XLQi,Ando}.

Recently, the layered, noncentrosymmetric compound PbTaSe$_2$ with heavy elements was found to be superconducting \cite{Cava}. It has a transition temperature $T_c$ = 3.72 K. Specific heat, electrical resistivity, and magnetic susceptibility measurements on PbTaSe$_{2}$ indicates a moderately coupled, type-II BCS superconductor. Importantly, PbTaSe$_2$ displays alternating stacking of hexagonal TaSe$_{2}$ and Pb layers which is noncentrosymmetric. The strong spin-orbit coupling (SOC) in PbTaSe$_2$ can lead to large Rashba splitting and the breaking of spin degeneracy \cite{Cava}. For such a noncentrosymmetric superconductor, there may exist nodes in the superconducting gap if the spin-triplet pairing state plays a significant role \cite{Bauer,Frigeri}. Therefore it will be very interesting to probe the superconducting gap structure of PbTaSe$_{2}$.

More interestingly, a stable topological nodal-line semimetal state was observed in the normal state of PbTaSe$_2$ by angle-resolved photoemission spectroscopy (ARPES) measurements \cite{Hasan}. In contrast to a Weyl semimetal whose conduction and valence bands cross each other at discrete points, the topological nodal-line semimetal has extended band touching along one dimensional curves in the Brillouin zone \cite{FangC,YuR,KimY,WengH}. The nodal lines in PbTaSe$_2$ are protected by a reflection symmetry of the space group \cite{Hasan}. For superconducting PbTaSe$_2$ with such topological band structure, more experiments are highly desired to determine whether it is a topological superconductor.

Ultra-low-temperature heat transport is an established bulk technique to study the superconducting gap structure \cite{Shakeripour}. The existence of a finite residual linear term $\kappa_0/T$ in zero magnetic field is an evidence for gap nodes \cite{Shakeripour}. The field dependence of $\kappa_0/T$ may further give support for a nodal superconducting state, and provide information on the gap anisotropy, or multiple gaps \cite{Shakeripour}.

In this Rapid Communication, we present the thermal conductivity measurements of PbTaSe$_{2}$ single crystal down to 80 mK ($T_c$/50). The negligible $\kappa_{0}/T$ in zero field and the ``$S$"-shape curve of field dependent $\kappa_{0}(H)/T$ strongly suggest multiple nodeless superconducting gaps in PbTaSe$_{2}$. We discuss the relation between this gap structure and its noncentrosymmetric crystal structure and topological band structure.

\begin{figure}
\includegraphics[clip,width=7.5cm]{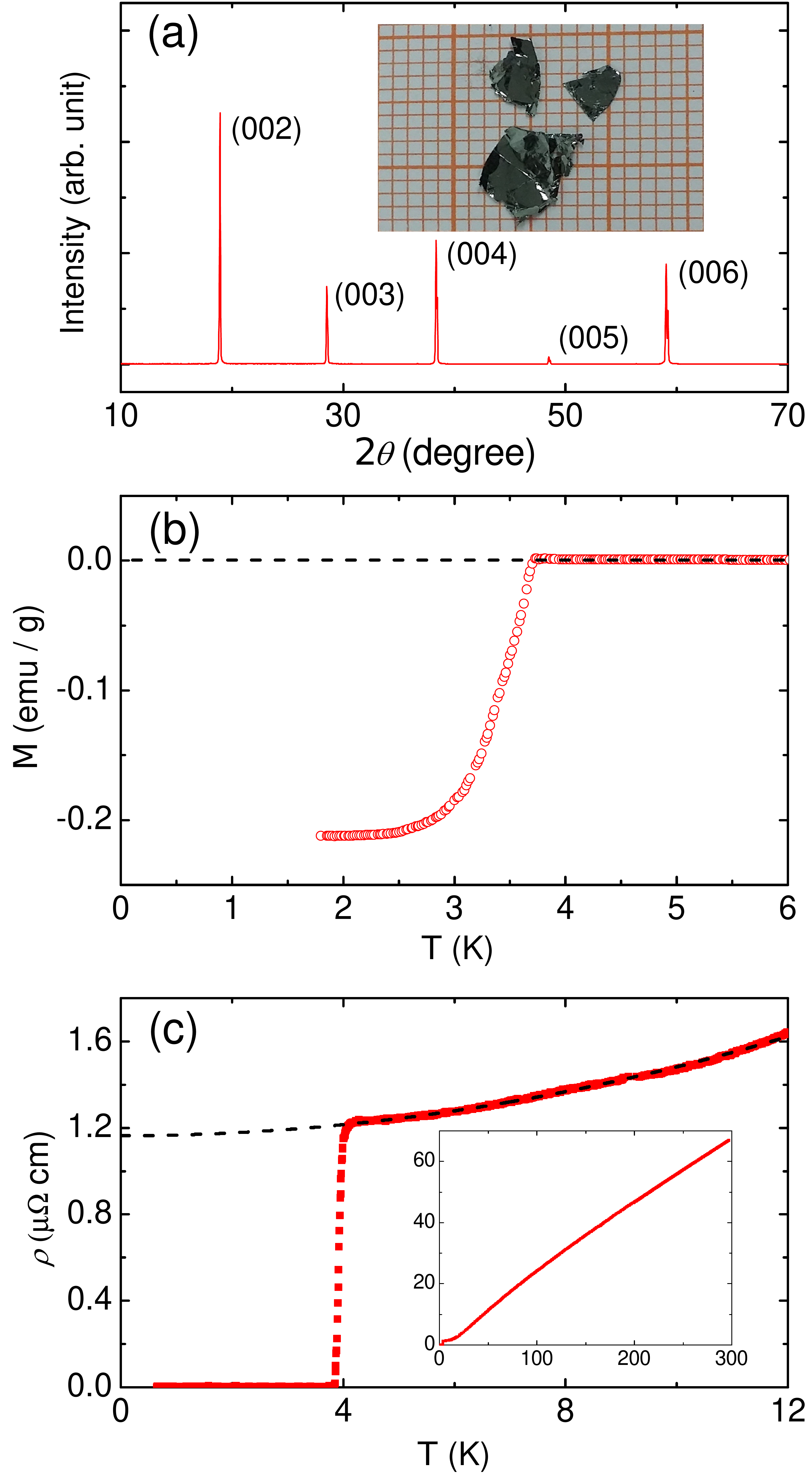}
\caption{(Color online). (a) X-ray diffraction pattern of PbTaSe$_{2}$ single crystal. Only (00$l$) Bragg peaks show up, demonstrating that the largest surface is $ab$ plan. Inset: Optical image of our PbTaSe$_{2}$ single crystals. (b) The dc magnetization at $H$ = 10 Oe for a PbTaSe$_{2}$ single crystal, measured with zero-field-cooled (ZFC) process. (c) The resistive superconducting transition of PbTaSe$_{2}$ single crystal at zero field. The dashed line is a fit of the normal-state resistivity between 4.5 and 12 K to $\rho$ = $\rho_0$ + $AT^2$. Inset: the temperature dependence of resistivity up to 297 K.}
\end{figure}

\begin{figure}
\includegraphics[clip,width=7.27cm]{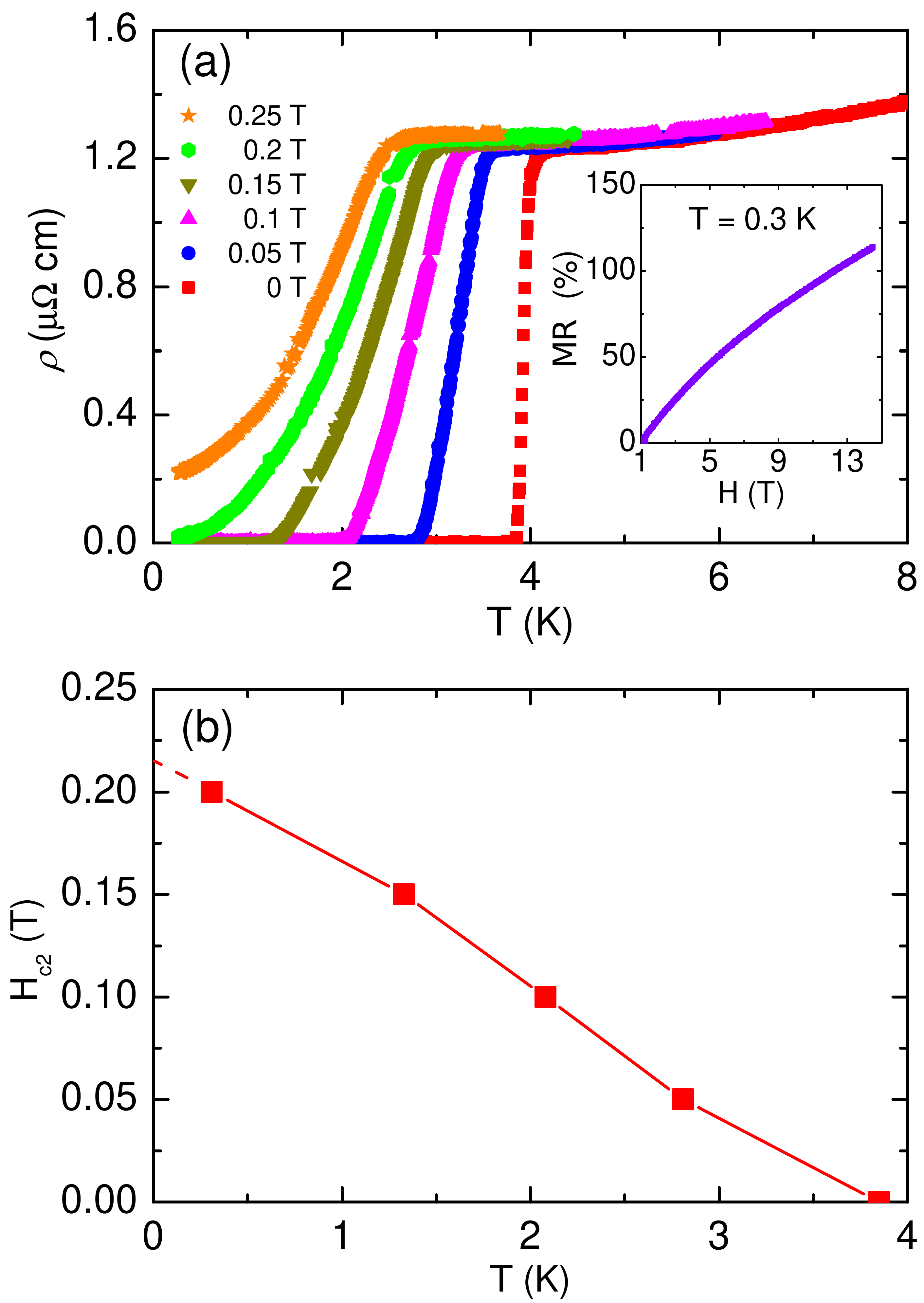}
\caption{(Color online). (a) Low-temperature resistivity of
PbTaSe$_{2}$ single crystal in magnetic field up to 0.25 T. Inset: the magnetoresistance of PbTaSe$_{2}$ single crystal at 0.3 K and above $H$ = 1 T. (b)
Temperature dependence of the upper critical field $H_{c2}(T)$,
defined by $\rho = 0$. The dashed line is a guide to the eye, which
points to $H_{c2}(0) \approx$ 0.22 T.}
\end{figure}

The PbTaSe$_{2}$ single crystals were grown by the chemical vapor transport (CVT) method as previously described in Ref. \cite{Hasan}. The obtained single crystals have typical size of 5 $\times$ 5 $\times$ 0.02 mm$^3$. The largest surface was identified as $ab$ plane by x-ray diffraction measurement, shown in Fig. 1(a). The dc magnetization measurements were performed in a superconducting quantum interference device (SQUID) [magnetic properties measurement system (MPMS), Quantum Design]. The sample for transport measurements was cut to a rectangular shape of 2.43 $\times$ 0.76 mm$^2$ in the $ab$ plane, with the thickness of 0.017 mm along $c$ axis. Four silver wires were attached to the sample with silver paint, which were used for both resistivity and thermal conductivity measurements, with electrical and heat currents in the (001) plane. The contacts are metallic with typical resistance of 15 m$\Omega$ at 2 K. The thermal conductivity was measured in a dilution refrigerator, using a standard four-wire steady-state method with two RuO$_2$ chip thermometers, calibrated {\it in situ} against a reference RuO$_2$ thermometer. Magnetic fields were applied perpendicular to $ab$ plane. To ensure a homogeneous field distribution in the sample, all fields for resistivity and thermal conductivity measurements were applied at temperature above $T_c$.

Figure 1(b) shows the dc magnetization of PbTaSe$_{2}$ single crystal. The onset of superconducting transition is at 3.74 K. The magnetization saturates below 2.5 K. The temperature dependence of resistivity at zero field is plotted in Fig. 1(c). From Fig. 1(c), the width of the resistive superconducting transition (10-90$\%$) is 0.12 K, and the $T_c$ defined by $\rho = 0$ is 3.84 K, which is consistent with the magnetization measurement. Fermi-liquid behavior $\rho \sim T^2$ is observed at low temperature in the normal state. The fit of $\rho(T)$ data between 4.5 and 12 K to $\rho$ = $\rho_0$ + $AT^2$ gives the
residual resistivity $\rho_{0}$ = 1.16 $\mu\Omega$ cm. The $\rho(T)$ curve up to 297 K is shown in the inset of Fig. 1(c), which gives the residual resistivity ratio (RRR) $\rho$(297 K)/$\rho_0 \approx$ 57.

Figure 2(a) plots the low-temperature resistivity of PbTaSe$_2$ single crystal in magnetic fields up to 0.25 T. The inset of Fig. 2(a) shows the magnetoresistance (MR) at 0.3 K and above $H$ = 1 T, which is defined as MR = [$\rho(H)-\rho$(1 T)]/$\rho$(1 T) $\times$ 100\%. The MR is quite significant in the normal state of PbTaSe$_2$. To estimate the upper critical field $H_{c2}$(0), the temperature dependence of $H_{c2}(T)$, defined by $\rho = 0$, is plotted in Fig. 2(b). $H_{c2}(0) \approx$ 0.22 T is roughly estimated.

The temperature dependence of thermal conductivity for PbTaSe$_{2}$ single crystal in zero and magnetic fields are plotted as $\kappa/T$ vs $T$ in Fig. 3. The measured thermal conductivity contains two contributions, $\kappa$ = $\kappa_e$ + $\kappa_p$, which come from electrons and phonons, respectively. In order to separate the two contributions, all the curves below 0.4 K are fitted to $\kappa/T$ = $a$ + $bT^{\alpha-1}$ \cite{MSutherland,SYLi}. The two terms $aT$ and $bT^{\alpha}$ represent contributions from electrons and phonons, respectively. The residual linear term $\kappa_0/T \equiv a$ is obtained by extrapolated $\kappa/T$ to $T$ = 0 K. Because of the specular reflections of phonons at the sample surfaces, the power $\alpha$ in the second term is typically between 2 and 3 \cite{MSutherland,SYLi}.

\begin{figure}
\includegraphics[clip,width=8.5cm]{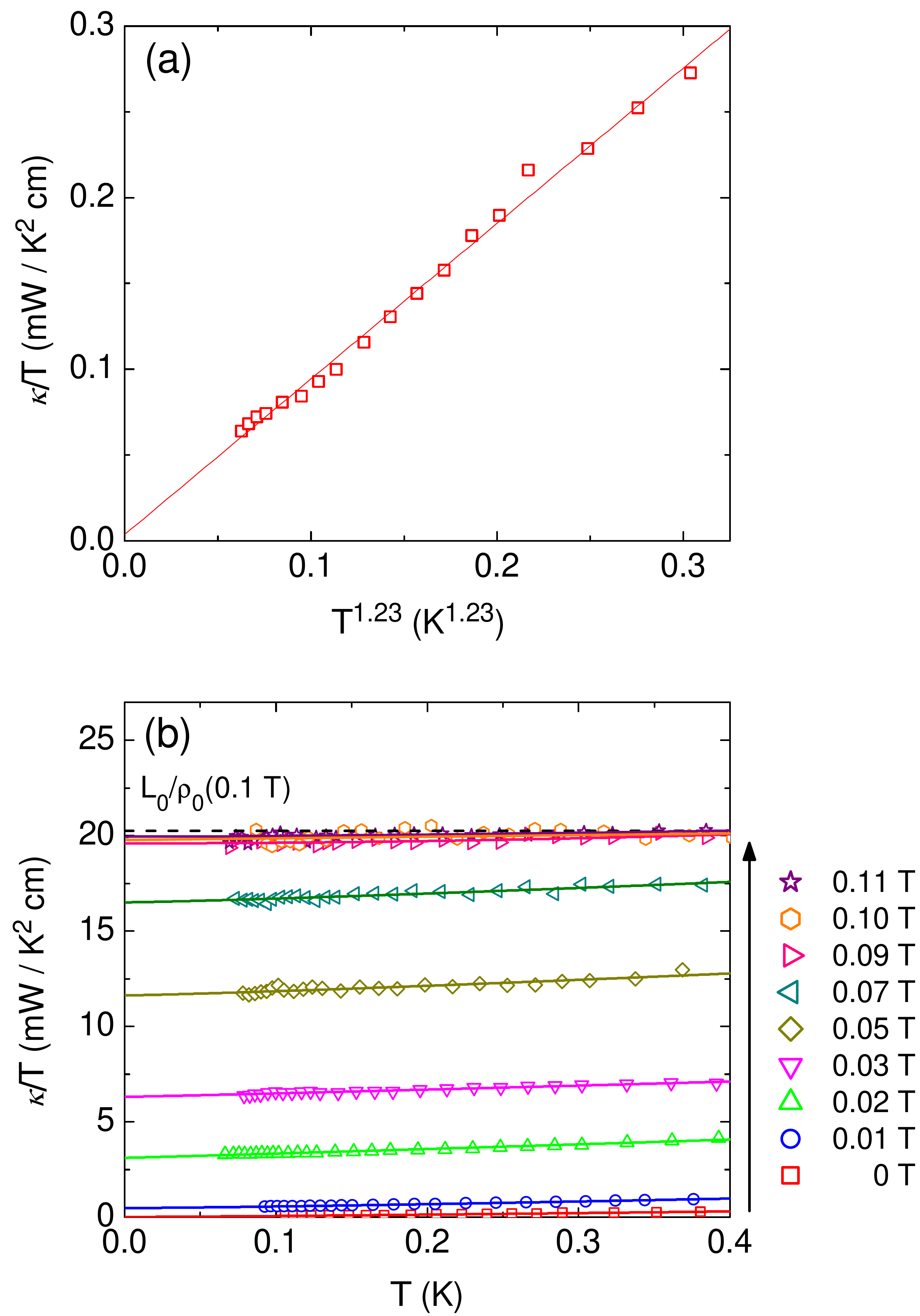}
\caption{(Color online). Low-temperature thermal conductivity
of PbTaSe$_2$ single crystals (a) in zero field, and (b) in magnetic fields
applied perpendicular to the $ab$ plane. The solid lines represent the fits to
$\kappa/T$ = $a$ + $bT^{\alpha-1}$ for the data in different $H$. The dashed lines are the
normal-state Wiedemann-Franz law expectation $L_0$/$\rho_0$(0.1 T), with
the Lorenz number $L_0 =$ 2.45 $\times$ 10$^{-8}$W $\Omega$ K$^{-2}$
and $\rho_0$(0.1 T) = 1.21 $\mu$$\Omega$ cm.}
\end{figure}

\begin{figure}
\includegraphics[clip,width=8.3cm]{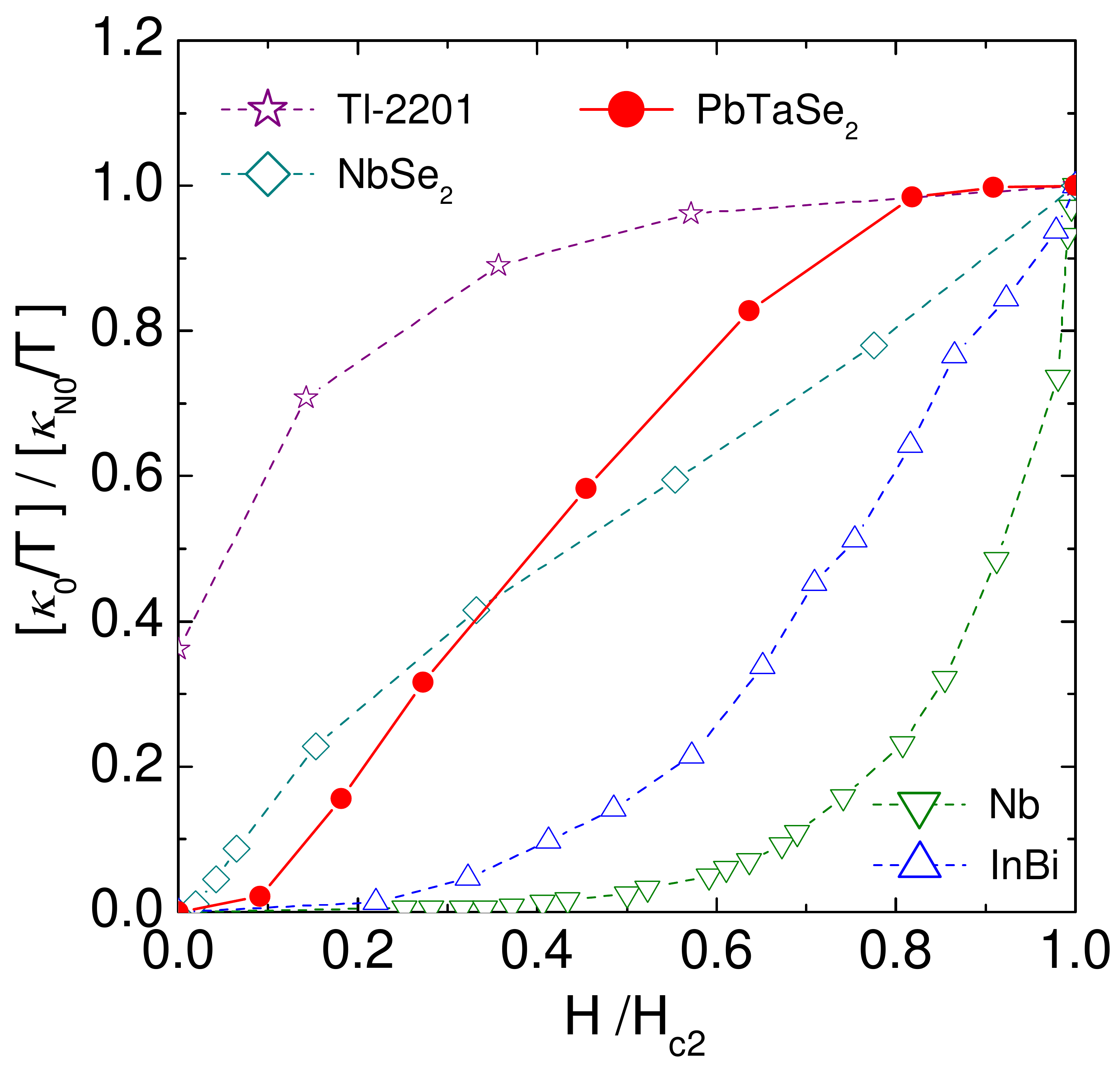}
\caption{(Color online). Normalized residual linear term
$\kappa_0/T$ of PbTaSe$_{2}$ as a function of $H/H_{c2}$, with bulk $H_{c2}$ = 0.11 T.
For comparison, similar data are shown for the clean $s$-wave
superconductor Nb \cite{Lowell}, the dirty $s$-wave superconducting
alloy InBi \cite{JOWillis}, the multiband $s$-wave superconductor
NbSe$_2$ \cite{EBoaknin}, and an overdoped $d$-wave cuprate
superconductor Tl-2201 \cite{Proust}.} \end{figure}

In zero field, the fitting gives $\kappa_0/T$ = 3 $\pm$ 10 $\mu$W K$^{-2}$ cm$^{-1}$ and $\alpha$ = 2.23. Comparing with our experimental error bar $\pm$ 5 $\mu$W K$^{-2}$ cm$^{-1}$, the $\kappa_0/T$ of PbTaSe$_2$ in zero field is negligible. For $s$-wave nodeless superconductors, there are no fermionic quasiparticles to conduct heat as $T \to 0$, since all electrons become Cooper pairs \cite{Shakeripour,MSutherland}. Therefore there is no residual linear term of $\kappa_0/T$, as seen in V$_3$Si and NbSe$_2$ \cite{MSutherland,EBoaknin}. However, for nodal superconductors, a substantial $\kappa_0/T$ in zero field contributed by the nodal quasiparticles has been found \cite{Shakeripour}. For example, $\kappa_0/T$ of the overdoped ($T_c$ = 15 K) $d$-wave cuprate superconductor Tl$_2$Ba$_2$CuO$_{6+\delta}$ (Tl-2201) is 1.41 mW K$^{-2}$ cm$^{-1}$, $\sim$36\% $\kappa_{N0}/T$ \cite{Proust}. For the $p$-wave superconductor Sr$_2$RuO$_4$ ($T_c$ = 1.5 K), $\kappa_0/T$ = 17 mW K$^{-2}$ cm$^{-1}$ was reported \cite{Suzuki}, more than 9\% $\kappa_{N0}/T$. Therefore, the negligible $\kappa_0/T$ of PbTaSe$_2$ strongly suggests that its superconducting gap is nodeless.

The field dependence of $\kappa_0/T$ can provide further information of the superconducting gap structure \cite{Shakeripour}. Between $H$ = 0 and 0.11 T, we fit all the curves and obtain the $\kappa_0/T$ for each magnetic field. In $H =$ 0.11 T, the fitting gives $\kappa_0/T =$ 19.95 $\pm$ 0.05 mW K$^{-2}$ cm$^{-1}$. This value of $\kappa_0/T$ roughly meets the normal-state Wiedemann-Franz law expectation $L_0$/$\rho_0$(0.1 T) = 20.2 mW K$^{-2}$ cm$^{-1}$, with the Lorenz number $L_0 =$ 2.45 $\times$ 10$^{-8}$W $\Omega$ K$^{-2}$ and $\rho_0$(0.1 T) = 1.21 $\mu$$\Omega$ cm. The verification of the Wiedemann-Franz law in the normal state demonstrates that our thermal conductivity measurements are reliable. We take $H =$ 0.11 T as the bulk $H_{c2}$ of PbTaSe$_2$. Note that this bulk $H_{c2}$ is only half of the $H_{c2}$ determined by resistivity measurements, which will be discussed later.

The normalized $\kappa_0(H)/T$ as a function of $H/H_{c2}$ for PbTaSe$_2$ is plotted in Fig. 4, with bulk $H_{c2}$ = 0.11 T. For comparison, similar data of the clean $s$-wave superconductor Nb \cite{Lowell}, the dirty $s$-wave superconducting alloy InBi \cite{JOWillis}, the multiband $s$-wave superconductor NbSe$_2$ \cite{EBoaknin}, and an overdoped $d$-wave cuprate superconductor Tl-2201 \cite{Proust}, are also plotted. For single band $s$-wave superconductor Nb, the $\kappa_0(H)/T$ changes little even up to 40\% $H_{c2}$ \cite{Lowell}. While for nodal superconductor Tl-2201, a small field can yield a quick growth in the quasiparticle density of states (DOS) due to Volovik effect, and the low field $\kappa_0(H)/T$ shows a roughly $\sqrt{H}$ dependant \cite{Proust}. In the case of NbSe$_2$, the distinct $\kappa_0(H)/T$ behavior was well explained by multiple superconducting gaps with different magnitudes \cite{EBoaknin}.

From Fig. 4, the curve of the normalized $\kappa_0(H)/T$ for PbTaSe$_2$ is close to that of the multiband $s$-wave superconductor NbSe$_2$. In fact, the curve is more like the ``$S$"-shape previously observed in nickel-based superconductors BaNi$_2$As$_2$ and TlNi$_2$Se$_2$ \cite{FRonning,XCHong}. Such an ``$S$"-shape $\kappa_0(H)/T$ curve is a strong evidence for multiple nodeless superconducting gaps \cite{XCHong}. In deed, the ARPES results clear show multiple bands in PbTaSe$_{2}$ \cite{Hasan}. This multiband superconductivity may explain the quite different $H_{c2}$ determined by resistivity and thermal conductivity measurements, imaging that most of the quasiparticles become normal near the bulk $H_{c2}$, but only a very small amount of carriers in the band with largest gap are still superconducting Cooper pairs.

Having demonstrated the superconducting gap structure of PbTaSe$_2$, we discuss its relation with the noncentrosymmetric crystal structure and topological band structure. So far, there are already many members in the family of noncentrosymmetric superconductors. However, only very few of them manifests unconventional superconductivity, such as CePt$_3$Si, CeRhSi$_3$, CeIrSi$_3$, and Li$_2$Pt$_3$B \cite{Bauer}. The absence of gap node in PbTaSe$_2$ indicates that the spin-singlet pairing state is dominant. In this sense, it may not be an unconventional superconductor.

From the aspect of topological superconductor, it requires fully gapped superconducting state. Our experimental result meets this requirement, if PbTaSe$_2$ with topological bulk nodal lines in the band structure is indeed a topological superconductor. However, to confirm this, the surface state of PbTaSe$_2$ needs to be carefully studied, for example by scanning tunneling microscopy (STM), to search for the Majorana fermions.

In summary, we have probed the superconducting gap structure of PbTaSe$_2$ by measuring the thermal conductivity down to 80 mK. The negligible $\kappa_0/T$ at zero field and the ``$S$"-shape field dependence of $\kappa_0/T$ demonstrate multiple nodeless superconducting gaps in PbTaSe$_2$. This result suggests that the spin-triplet state with gap nodes does not play an important role in this noncentrosymmetric superconductor. Further works are needed to investigate whether the superconducting state of PbTaSe$_2$ is topological or not.

This work is supported by the Ministry of Science and Technology of China (National Basic
Research Program No: 2012CB821402 and 2015CB921401), the Natural Science Foundation of China, the Program
for Professor of Special Appointment (Eastern Scholar) at Shanghai Institutions of Higher Learning, and STCSM of China (No. 15XD1500200). \\

$^*$ E-mail: shiyan$\_$li@fudan.edu.cn

\end{document}